\documentstyle[12pt]{article}
\renewcommand{\baselinestretch}{1.3}
\begin{document}
\title{Weak Hyperon Decays: Quark Sea and SU(3) Symmetry Breaking}
\author{P. \.{Z}enczykowski\\
\\
Dept. of Theor. Physics\\
H. Niewodnicza\'nski Institute of Nuclear Physics\\
Radzikowskiego 152, Krak\'ow, Poland\\
}
\maketitle
\begin{abstract}
An explanation of the difference in the values of the apparent $f/d$ ratios for
the S- and P- wave amplitudes of nonleptonic hyperon decays is proposed. The
argument is formulated in the framework of the standard pole model with
$(56,0^{+})$ ground-state and $(70,1^{-})$ excited baryons as intermediate
states for the P- and S- waves respectively. Under the assumption that the
dominant part of the deviation of $(f/d)_{P-wave}$ from $-1$ is due to large
quark sea effects, $SU(3)$ symmetry breaking in energy denominators is
shown to lead to a prediction for $(f/d)_{S-wave}$ which is in excellent
agreement with experiment. This corroborates our previous unitarity
calculations which indicated that the matrix elements $<B|H^{p.c.}_{weak}|B'>$
of the parity conserving weak Hamiltonian between the ground-state baryons are
characterized by $f_{0}/d_{0} \approx -1.6$ or more. A brief discussion of the
problem of the relative size of S- and P- wave amplitudes is given.
Finally, implications for weak radiative hyperon decays
are also discussed.
\end{abstract}

\newpage
\section{Introduction}
Despite several decades of theoretical inquiry, our understanding of weak
hyperon decays has remained elusive and controversial \cite{DGH86}. Dominantly,
hyperons decay weakly into two-body pion+baryon channels. Various models
proposed for a theoretical description of these nonleptonic processes always
relate to an approach based on PCAC and current algebra (CA) \cite{Gel64}. One
of the reasons for such a pronounced role of that approach is that it is
theoretically attractive: it allows a parallel treatment of the S- and P-waves,
expressing both of these as functions of the transition matrix elements
$<B'|H^{p.c.}_{weak}|B>$ of the
parity conserving part of the weak Hamiltonian.

Unfortunately, this PCAC/CA approach is less appealing when confronted with
experiment as it presents us with two serious difficulties.
The first concerns the relative size of the S- and P- waves: current algebra
overestimates the S:P ratio by a factor of around 2. The second is related to
the SU(3) structure of the decays. The quark model prediction for the two
$SU(3)$-invariant couplings $f_{0},d_{0}$ describing the $SU(3)$ structure of
the $<B'|H^{p.c.}_{weak}|B>$ matrix elements is $f_{0}/d_{0}=-1$, while the
experimental S-waves require $f/d \approx -2.5$. Similarly, the value of the
$f/d$ ratio extracted from the P-waves is different from -1. Its exact value is
sensitive to the way one treats $SU(3)$ breaking in energy denominators and
couplings. When $SU(3)$-symmetric $\pi BB'$ couplings and equal spacing of
ground-state octet baryons are used one infers from the P-wave amplitudes that
$f/d \approx -1.8$ or $-1.9$ \cite{DGH86,Zen89,Zen91}

As yet there is no general consensus as to what a full resolution of the above
problems might be. On one side, it is rather generally acknowledged that an
important correction to the CA results stems from a more realistic treatment
of the contribution
from the intermediate
$(70,1^{-})$ baryons.
In particular, SU(3) breaking in energy denominators generates
corrections which
subtract
from the standard soft pion contribution ~\cite{LeY79} . The correction is of
order $\delta
s / \Delta \omega _{s} \approx 0.3$ to $0.4$ relative to that of the commutator
( $\delta s$ is the $SU(3)$ breaking parameter ($\approx$ 190 MeV) and
$\Delta \omega _{s}$ is the mean spacing of $(56,0^{+})$ and $(70,1^{-})$
baryons). On the other side, however, no such consensus has been reached so far
on the question of the $f/d$ ratio. In fact, several different explanations of
the deviation of $f/d$ from $-1$ have been proposed.

In their original paper \cite{LeY79} LeYaouanc et al. have suggested that $f/d$
is larger in parity violating amplitudes because for different decays
such as $\Lambda
\rightarrow N \pi , \Sigma \rightarrow N \pi$,... the corrections due to
$(70,1^{-})$ baryons appear to be proportional to different mass differences of
ground-state baryons ($\Lambda - N, \Sigma - N, ...$). With $\Sigma - \Lambda
\neq 0$ one obtains then an increase of the {\em effective} $f/d$ ratio. The
problem
with this explanation is that $\Sigma - \Lambda$ splitting is a second order
effect due to spin-spin interactions which were neglected in the intermediate
$(70,1^{-})$ baryons in ref.~\cite{LeY79}. If spin-spin interactions are also
neglected for ground-state baryons one recovers for the $(70,1^{-})$ correction
the canonical quark model value $f/d = -1$.

Another possible and at first sight natural explanation is to attribute the
departure of $f/d$ from $-1$ to a contribution of diagrams with weak
Hamiltonian acting in the meson leg. Such diagrams are characterised by
$d_{mes}/f_{mes} = 0$ and thus they might provide the much needed
enhancement of $f$. For the S-waves they were invoked by Gronau \cite{Gro72}
who introduced the contribution of $K^{*}$ intermediate meson. The
contribution of such diagrams has been later discussed in various papers by
Bonvin \cite{Bon84}, Nardulli \cite{Nar88}, Xu and Stech \cite{XS91}, and
others. The main problem with this line of reasoning is that one expects such
contributions to be small on general grounds. Indeed, for the P-waves the
K-pole contribution  is proportional to
$p_{\pi} \cdot p_{K} \sim m^{2}_{\pi }$ as a result of chiral symmetry
(ref.~\cite{DGH86}) and it should vanish for $m^{2}_{\pi} \rightarrow 0$.
For the S-waves one can show that in the limit of exact $SU(3)$
symmetry such diagrams should give a vanishing contribution as well
(see e.g. ref.~\cite{DDH80}). In the case
of broken $SU(3)$ one might expect corrections to the quark model value of $-1$
of order $\delta s / (hadron~ mass~ scale) \approx 20-30\%$ but not $100-150\%$
!

The third possibility discussed in the literature consists in a large departure
from the assignment of the canonical value of $f_{0}/d_{0} = -1$ to the
(directly not measurable) matrix elements $<B|H^{p.c.}_{weak}|B'>$ of the
parity conserving part of the
weak Hamiltonian between the ground-state baryons.
This departure is atributed to the contribution from the sea quarks
\cite{DDH80,DG77}. In quantum chromodynamics this corresponds to the
consideration of penguin diagrams. On one side, direct evaluation of these
diagrams leads to
a small increase of $f/d$ only \cite{DGPH}. On the other side, if
one estimates the contribution of the penguins by relating them to the
gluon-induced $\Delta - N$ splitting one obtains \cite{DG77} a substantial
increase of
$f_{0}/d_{0}$ to $-1.6$ .
Although the size of this renormalization of $f/d$ is determined by the
experimentally observed $\Delta - N $ splitting, it corresponds to a large
value of the QCD coupling constant, believed by many to be unrealistic (see,
however, ref.~\cite{Eeg85}). A different origin for a large contribution from
sea quarks has been proposed recently in ref.~\cite{Zen92}. It has been shown
there that the interference of strong and parity-conserving weak (P-wave)
amplitudes leads to a substantial increase of the $f_{0}/d_{0}$ ratio
characterising the $<B'|H^{p.c.}_{weak}|B>$ matrix elements. When the size of
hadronic loops thus generated by unitarity is estimated by comparison with
hadron mass splittings one finds that $f_{0}/d_{0}$ is shifted by such
hadronic penguins to around $-1.6$ or more. The exact
value depends slightly on how much of the $\Delta - N$ splitting is attributed
to hadron-level (unitarity) effects. Even with moderate (around 80 MeV)
pion-induced contribution to the $\Delta - N$ splitting one obtains
$f_{0}/d_{0} = -1.5$ (ref.~\cite{Zen92}). For larger contributions of this type
as in the unitarised quark model \cite{Tor,Zen86} one gets $f_{0}/d_{0}$
around -1.6 or more. Thus, one can have both a smaller QCD coupling governing
the short distance effects and large (hadron-level induced) sea effects.

In this paper we study in more detail how these sea effects manifest themselves
in S- and P-wave amplitudes. We work in the framework of a kind of "skeleton"
pole model which both includes the essential $SU(3)$ breaking effects of the
pole model and - at the same time - retains much of the simplicity of the
PCAC/CA approach by  bypassing the need to use a detailed information on the
$\frac{1}{2} ^{-}$ baryons in the intermediate states.

We find that the model thus constructed explains the $f/d$ structure of both
the P- and S- wave amplitudes very naturally. In fact, joint consideration of
large quark sea effects and SU(3) breaking in energy denominators leads,
{\em without any new parameters}, to the following approximate relationship
between the deviations from $-1$ of the observed \footnote[1]{if meson-leg
contributions to $f$ are small} $f/d$ ratios in S- and P-wave amplitudes:
\begin{equation}
\label{eq:in1}
\frac{(f/d+1)_{S-wave}}{(f/d+1)_{P-wave}} = \frac{1+x}{1-x}
\end{equation}
where $x=\frac{\delta s}{ \Delta \omega_{s}} \approx 0.3$ to $0.4$.

Using the experimental values for the corresponding $f/d$ ratios (-2.6 for
S-waves, -1.85 to -1.9 for P-waves), Eq.(\ref{eq:in1}) reads: $1.8$~to~$1.9~=
2.1~\pm~0.25$. The experimentally observed deviation of $(f/d)_{P-wave}$ from
$-1$ is in agreement with the unitarity-based calculation \cite{Zen92} of the
$SU(3)$ structure of the $<B'|H^{p.c.}_{weak}|B>$ matrix elements:
$(f/d)_{P-wave}~\approx f_{0}/d_{0}$, or -1.8 to -1.9 $\approx$ -1.6 to -1.7.
This is consistent with general hadron level arguments permitting only a small
correction from meson-leg diagrams to $(f/d)_{P-wave}$.

The paper is organized as follows. In the next Section we exhibit the basic
$SU(3)$-symmetric connections between the quark diagrams, the pole model and
the PCAC/CA approach for the S-wave amplitudes. In Section 3 standard
description of the P-wave amplitudes and the assignment of the dominant part of
the deviation of $(f/d)_{P-wave}$ from $-1$ to quark sea effects is discussed
in
some detail. Section 4 contains the analysis of the $SU(3)$-symmetry breaking
effects in the energy denominators of the pole model for the S-wave amplitudes.
Eq.(\ref{eq:in1}) is derived there. It is also shown there that the S-wave
reduction mechanism of LeYaouanc et al. becomes unimportant for $f_{0}/d_{0}
\approx -1.7$. In an attempt to deal with this reappearing S:P problem , in
Section 5 we briefly consider the contribution from the radially excited
$(56,0^{+})^{*} \frac{1}{2}^{+}$ baryons. We find that, if the relevant
$f^{*}/d^{*}$ ratio is equal to that of ground-state baryons, the contribution
of radially excited states cannot cure the S:P problem. We argue then that the
smallness of the experimental S:P ratio may be related to the departure of the
ratio
$g_{B(\frac{1}{2}^{+})B^{*}(\frac{1}{2}^{-})P}/
g_{B(\frac{1}{2}^{+})B'(\frac{1}{2}^{+})P}$
of strong hadron couplings from quark model predictions. In Section 6 a brief
discussion is given of the modifications to the combined symmetry - vector
meson dominance approach to weak radiative hyperon decays, that originate from
the effect considered in this paper. Finally, in Section 7 we reiterate the
main points of our paper.

\section{The Parity Violating Amplitudes}
\label{pv}
All quark-line diagrams that may in principle contribute to weak hyperon
decays are shown in Fig.1. Diagrams $(a)$ and $(a')$ correspond to the
meson-leg topology, while diagrams $(b)$, $(c)$, $(d)$ and $(e)$ admit
intermediate
baryons in between the action of the weak Hamiltonian and the strong
(meson-emission) vertex.

For the parity violating amplitudes the contributions from diagrams $(a)$,
$(a')$ vanish in the $SU(3)$-symmetry limit \cite{DDH80}. Similarly, Lee-Swift
theorem
\cite{LeeSw} requires the vanishing of diagrams $(d)$ and $(e)$. Diagrams $(b)$
are the familiar $W$-exchange processes that lead to $f/d=-1$, while diagrams
$(c)$ are the sea diagrams (with $d=0$). In an $SU(6)_{W}$ symmetric approach
the contributions from the diagrams $(b1)$, $(b2)$, $(c1)$, and $(c2)$ can be
calculated using the quark model technique of Desplanques, Donoghue and
Holstein \cite{DDH80} and are gathered in Table 1. For completeness the weights
for the kinematically forbidden transitions are also given.

In terms of the reduced matrix elements $b$ and $c$ corresponding to diagrams
$(b1)$, $(b2)$ and $(c1)$, $(c2)$ respectively one obtains from Table 1 the
following expressions for the parity violating amplitudes:
\begin{eqnarray}
\label{eq:pv2}
A(\Sigma ^{+}_{0}) & = &\frac{1}{2\sqrt{2}}~b -
\frac{1}{6\sqrt{2}}~ c \nonumber \\
A(\Sigma ^{+}_{+}) & = & 0 \nonumber \\
A(\Sigma ^{-}_{-}) & = & -\frac{1}{2}~ b + \frac{1}{6}~ c
\nonumber \\ A(\Lambda ^{0}_{-})~ =~  -\sqrt{2} A(\Lambda ^{0}_{0}) & = &
-\frac{1}{2\sqrt{6}}~ b + \frac{1}{2\sqrt{6}}~ c \nonumber \\
A(\Xi ^{-}_{-})~ =~ -\sqrt{2} A(\Xi ^{0}_{0}) & = & \frac{1}{\sqrt{6}}~ b -
\frac{1}{2\sqrt{6}}~ c
\end{eqnarray}
For the kinematically forbidden amplitudes one gets similarly
\begin{equation}
\label{eq:pv2a}
A(\Sigma ^{+} \rightarrow p \eta _{8}) =
\left( -\frac{1}{\sqrt{6}}-\frac{1}{2\sqrt{6}} \right) ~b +
\left( \frac{1}{6\sqrt{6}}+\frac{1}{3\sqrt{6}} \right) ~c
\end{equation}
etc. (i.e. the entries from Table 1 that correspond to diagrams $(b1)$, $(b2)$
($(c1)$, $(c2)$) are to be {\em added}). Experiment fixes then $b=-5$, $c=+12$
(in units of $10^{-7}$, see ref.~\cite{Zen89}), $f/d = -1+(2c)/(3b) \approx
-2.6$.

Let us discuss how formulas (\ref{eq:pv2}) are related to the pole model and
the PCAC/CA approach. For the sake of definiteness consider the $\Sigma ^{+}
\rightarrow p \pi ^{0}$ decay. Upon using the PCAC relation between the pion
field and the divergence of the axial current, the calculation of the S-wave
amplitude $A(\Sigma ^{+} \rightarrow p \pi ^{0})$ in the pole model involves
the consideration of the expressions
\begin{equation}
\label{eq:pv3a}
A_{(1)}(\Sigma ^{+} \rightarrow p \pi ^{0})  =
\frac{<p|\partial _{\mu}A_{\mu}^{(0)}|N^{*}><N^{*}|H^{p.v.}_{weak}|\Sigma
^{+}>}{\Delta
\omega _{W1}}
\end{equation}
and
\begin{equation}
\label{eq:pv3}
A_{(2)}(\Sigma ^{+} \rightarrow p \pi ^{0})  =
\frac{<p|H^{p.v.}_{weak}|\Sigma ^{*}><\Sigma ^{*}|\partial
_{\mu}A_{\mu}^{(0)}|\Sigma
^{+}>}{\Delta
\omega _{W2}}
\end{equation}
corresponding to diagrams $(b1), (c1)$ and $(b2), (c2)$ respectively. (We have
ignored uninteresting factors such as $1/f_{\pi}$ on the r.h.s. of
Eqs.(\ref{eq:pv3a},\ref{eq:pv3})). In Eqs.(\ref{eq:pv3a},\ref{eq:pv3}) the
dominant contribution is expected
to arise from the $N^{*}$ and $\Sigma ^{*}$ $(70,1^{-})~\frac{1}{2}^{-}$
intermediate states. The energy denominators $\Delta \omega _{W1}$, $\Delta
\omega _{W2}$ have subscripts $_{W1}$ ($_{W2}$) since they
correspond to energy difference "across" the {\em weak} interaction:
\begin{eqnarray}
\label{eq:pv3c}
\Delta \omega _{W1} & = & N^{*} - \Sigma \nonumber \\
\Delta \omega _{W2} & = & \Sigma^{*} - p
\end{eqnarray}

Since the matrix elements of the spatial components
$A_{k}$ of the
axial current between $<p|$ and $|N^{*}>$ ($<\Sigma ^{*}|$ and $|\Sigma ^{+}>$)
vanish (see ref.~\cite{LeY79}), we have
\begin{eqnarray}
\label{eq:pv4}
\frac{1}{i} <p~|\partial _{\mu}A_{\mu}|N^{*}> & = & ~\Delta \omega _{s}
<p~|A_{0}|N^{*}> \nonumber \\
\frac{1}{i} <\Sigma^{*}|\partial _{\mu}A_{\mu}|\Sigma^{+}> & = & - \Delta
\omega _{s}
<\Sigma ^{*}|A_{0}|\Sigma ^{+}>
\end{eqnarray}
where we have used the subscript $_{s}$ to denote
baryon energy difference "across" the {\em strong}~ vertex
\begin{equation}
\label{eq:pv4c}
\Delta \omega _{s} = N^{*} - p = \Sigma ^{*} - \Sigma
\end{equation}
In the $SU(3)$ limit we have $\Delta \omega _{s} = \Delta \omega _{W1} = \Delta
\omega _{W2}$ and one obtains from Eqs.(\ref{eq:pv3a},\ref{eq:pv3}):
\begin{eqnarray}
\label{eq:pv5}
A_{(1)}(\Sigma ^{+} \rightarrow p \pi ^{0}) & = &
{}~<p~|A_{0}|N^{*}><N^{*}|H^{p.v.}_{weak}|\Sigma
^{+}> \nonumber \\
A_{(2)}(\Sigma ^{+} \rightarrow p \pi ^{0}) & = &
- <p~|H^{p.v.}_{weak}|\Sigma ^{*}><\Sigma ^{*}|A_{0}|\Sigma ^{+}>
\end{eqnarray}
i.e. we recover the standard commutator prescription of current algebra:
\begin{equation}
\label{eq:pv6}
A = A_{(1)} + A_{(2)} = <p~|~[A_{0},H^{p.v.}_{weak}]~|\Sigma^{+}>
\end{equation}
which, upon using the commutation relation
\begin{equation}
\label{eq:pv7}
[A_{0},H^{p.v.}_{weak}] = [V_{0},H^{p.c.}_{weak}]
\end{equation}
enables us to express $A(\Sigma ^{+}_{0})$ in terms of the matrix element
$<p|H^{p.c.}_{weak}|\Sigma ^{+}>$.

\section{The Parity Conserving Amplitudes}
\label{pc}
The $SU(6)$ structure of the parity conserving amplitudes corresponding to the
diagrams of Fig.1 may be calculated using, as before, the quark model technique
of refs.~\cite{DDH80,Zen89}. This time, however, the dominant contribution is
expected to come from the ground-state baryons as intermediate states. This
introduces energy denominators (here for $Y \rightarrow N\pi$ processes)
\begin{equation}
\label{eq:pc8}
\frac{1}{N-Y} ~~~~~\left( \frac{1}{Y-N} \right)
\end{equation}
for diagrams $(b1)$, $(c1)$ ($(b2)$, $(c2)$) respectively. On account of the
sign difference between these energy denominators the $SU(6)$ factors
corresponding to diagrams $(b1)$, $(b2)$ should be subtracted (and similarly
for diagrams $(c1)$, $(c2)$). For diagrams $(d1)$, $(d2)$ and $(e1)$, $(e2)$
this subtraction procedure leads to the total cancellation of their
contributions. The relevant $SU(6)$ factors are gathered in Table 2, where, for
completeness, the factors corresponding to the separate diagrams $(b1)$,
$(b2)$, $(c1)$, and $(c2)$ are given. Contributions from the individual
diagrams $(d1)$, $(d2)$, $(e1)$, and $(e2)$ - though nonzero in general - are
not shown. The entries in Table 2 correspond to the $F/D$ ratio of $SU(6)$,
i.e. equal to $2/3$. Phenomenologically more successful fits are obtained in
the pole models in which $F/D$ differs slightly from its $SU(6)$ value: $F/D
\approx 0.56$ or $0.58$. Explicit dependence on $F/D$ of the ground-state
baryon pole model formulas is given in Eq.(\ref{eq:pc9}). (See also Table 3
where weights of individual baryon pole contributions corresponding to the two
($(1)$ and $(2)$) different orderings of the strong and weak transitions (see
Fig.2) are exhibited. Table 3 includes the effects of all the quark diagrams
$(b)$, $(c)$, $(d)$, and $(e)$.)

In Eq.(\ref{eq:pc9}) $f_{0}/d_{0}$ characterizes the $<B'|H^{p.c.}_{weak}|B>$
matrix elements, while all energy denominators $\pm \frac{1}{N-Y}$ (we use
$\Sigma - N = \Lambda - N = \Xi - \Sigma$) are contained in the overall
normalization factor $C~=~-33$ (see also ref.\cite{Zen89}):
\begin{eqnarray}
\label{eq:pc9}
B(\Sigma ^{+}_{0}) & = &
\frac{1}{\sqrt{2}}\left( \frac{f_{0}}{d_{0}}-1\right) \left( 1-\frac{F}{D}
\right) C
\nonumber
\\ B(\Sigma ^{+}_{+}) & = & -\frac{4}{3} C \nonumber \\
B(\Sigma ^{-}_{-})  & = & \left[ \left( \frac{f_{0}}{d_{0}}-1 \right)
\frac{F}{D} -\frac{1}{3}
\left( 3\frac{f_{0}}{d_{0}}+1 \right) \right] C \nonumber \\
B(\Lambda^{0}_{-}) = -\sqrt{2} B(\Lambda^{0}_{0}) & = & \frac{1}{\sqrt{6}}
\left[ \frac{f_{0}}{d_{0}}+3+\left( 3\frac{f_{0}}{d_{0}}+1 \right) \frac{F}{D}
\right] C
\nonumber \\
B(\Xi^{-}_{-}) = -\sqrt{2} B(\Xi^{0}_{0}) & = & -\frac{1}{\sqrt{6}}
\left[ 3-\frac{f_{0}}{d_{0}}+ \left( 3\frac{f_{0}}{d_{0}}-1 \right)
\frac{F}{D} \right] C
\end{eqnarray}
The correspondence between the expressions resulting from the use of Table~2
through
\begin{equation}
\label{eq:pc10a}
B(\Sigma ^{+}_{0}) = -\frac{1}{6\sqrt{2}}~\beta - \frac{1}{9\sqrt{2}}~\gamma
\end{equation}
and Table 3 through Eq.(\ref{eq:pc9}) is given by taking in Eq.(\ref{eq:pc9})
$F/D=2/3$ and identifying
\begin{eqnarray}
\label{eq:pc10b}
\beta & = & 4 C \nonumber \\
\gamma  & = & -3 \left( 1+\frac{f_{0}}{d_{0}} \right) C
\end{eqnarray}
In Eq.(\ref{eq:pc10a})~ $\beta$ and $\gamma$ are the reduced matrix elements
corresponding to diagrams $(b1)$, $(b2)$ and $(c1)$, $(c2)$ respectively. As is
clearly seen from Eq.(\ref{eq:pc10b}), in the ground-state baryon pole model
the deviation of the experimentally observed $f/d$ from its canonical value of
$-1$ is attributed to a substantial contribution from diagrams $(c)$ which
modifies the $f_{0}/d_{0}$ structure of the $<B'|H^{p.c.}_{weak}|B>$ matrix
elements. Eq.(\ref{eq:pc9}) describes the P-wave data very well (see Table 4).
Note that one cannot expect here a better agreement in view of the violation of
the $\Delta I = 1/2$ rules by the data. For example, the $\Delta I = 1/2$ rule
$\sqrt{2} \Sigma ^{+}_{0} = \Sigma ^{+}_{+} - \Sigma ^{-}_{-}$ experimentally
reads $37.6 \pm 1.8 = 43.8 \pm 0.4$. The not-well-understood $\Delta I = 3/2$
amplitudes are of the order of a few percent.

{}From Table 4 we see that the data seem to require $(f/d)_{P-wave} \approx
-1.85$ to $-1.9$. The ground-state baryon pole model identifies this $f/d$ as
the $f_{0}/d_{0}$ ratio characterizing the $<B'|H^{p.c.}_{weak}|B>$ matrix
elements. Although it is hard to make a fully reliable calculation of sea quark
effects, the estimates of $f_{0}/d_{0}$ performed by Donoghue and Golowich
\cite{DG77} and by the author \cite{Zen92} lead to $f_{0}/d_{0} \approx -1.6$
or more. In ref.\cite{DG77} quark sea effects are due to short distance QCD
interactions while in ref.\cite{Zen92} hadron-level unitarity plays the
dominant role in boosting the value of $f_{0}/d_{0}$ away from $-1$. The
precise division of how much of this shift is due to short and long distance
effects is not important here: the size of the quark sea contribution is
determined by the total of these effects. It is this total that can be directly
linked with the experimental value of the $\Delta - N$ splitting. In this way
large deviation of $f_{0}/d_{0}$ from $-1$ is correlated with the size of the
$\Delta - N$ splitting.

In conclusion, it is natural to expect that the dominant part of the deviation
of $(f/d)_{P-wave}$ from $-1$ is due to quark sea effects as identified in
Eq.(\ref{eq:pc10b}) and that $f_{0}/d_{0}$ is close to (say) $-1.7$.
The remaining small enhancement of $(f/d)_{P-wave}$ may come from the meson-leg
diagrams. For example, Xu and Stech \cite{XS91} estimate the
contribution to $f$ arising
from non-penguin factorization diagrams to be around
$f_{non-penguin}/d_{0}
\approx -0.15$ to $-0.2$.

\section{Quark Sea Effects in S- Waves}
\label{sea}
To reconcile the value of the $f/d$ ratio observed in the P-wave amplitudes
with the one needed for a proper description of the S-wave amplitudes we shall
consider $SU(3)$ symmetry breaking in the energy denominators of the latter.
This effect was originally discussed by LeYaouanc et al. \cite{LeY79} who have
shown how its inclusion works towards reducing the discrepancy in size between
the CA estimate of the S:P ratio and experiment. What LeYaouanc et al. did not
consider was the presence of $SU(3)$ symmetry breaking in denominators {\em in
conjunction} with large quark sea effects. When $SU(3)$ breaking is taken into
account, the r.h.s. of Eqs.(\ref{eq:pv5}) are modified and one obtains
\begin{eqnarray}
\label{eq:sea11}
A_{(1)}(\Sigma ^{+} \rightarrow p \pi ^{0}) & = &
\frac{\Delta \omega_{s}}{\Delta \omega_{s} -\delta s}
<p|A_{0}|N^{*}><N^{*}|H^{p.v.}_{weak}|\Sigma
^{+}> \nonumber \\
A_{(2)}(\Sigma ^{+} \rightarrow p \pi ^{0}) & = &
- \frac{\Delta \omega_{s}}{\Delta \omega_{s} +\delta s}
<p|H^{p.v.}_{weak}|\Sigma ^{*}><\Sigma ^{*}|A_{0}|\Sigma ^{+}>
\end{eqnarray}
In Eqs.(\ref{eq:sea11}) we have put
\begin{eqnarray}
\label{eq:sea11a}
\Delta \omega _{W1} = N^{*}-\Sigma & = & \Delta \omega_{s} - \delta s \nonumber
\\
\Delta \omega _{W2} = \Sigma^{*}-p & = & \Delta \omega_{s} + \delta s
\end{eqnarray}
 with $\Delta \omega _{s} \approx 570 MeV$ being the average splitting between
the $(56,0^{+}) \frac{1}{2} ^{+}$ and $(70,1^{-})\frac{1}{2} ^{-}$ multiplets
and
$\delta s
\approx 190 ~MeV$ being the mass difference associated with a change of
strangeness by $-1$.
The sums over intermediate states $N^{*}$, $\Sigma^{*}$ on the r.h.s. of
Eqs.(\ref{eq:sea11}) are implicit in the weights of Table 1. These weights are
in turn proportional to the numerators of the pole model amplitudes. Using
Table 1, the sums in Eqs.(\ref{eq:sea11}) may be expressed therefore as
\begin{eqnarray}
\label{eq:sea12}
<p|A_{0}|N^{*}><N^{*}|H^{p.v.}_{weak}|\Sigma
^{+}> & = & -\frac{1}{6\sqrt{2}}~ k~ c_{0}\nonumber \\
-<p|H^{p.v.}_{weak}|\Sigma ^{*}><\Sigma ^{*}|A_{0}|\Sigma ^{+}> & = &
+\frac{1}{2\sqrt{2}}~ k~ b_{0}
\end{eqnarray}
where $k$ is some proportionality constant and $b_{0}$, $c_{0}$ are the $SU(3)$
invariant couplings characterising the $<B|H^{p.c.}_{weak}|B'>$ matrix
elements. Indeed, if there is no $SU(3)$ breaking in energy denominators, from
Eqs.(\ref{eq:sea11}) and (\ref{eq:sea12}) we obtain the $A(\Sigma ^{+}_{0})$
amplitude (which is proportional to the $<p|H^{p.c.}_{weak}|\Sigma ^{+}>$
matrix element) of Eq.(\ref{eq:pv2}) with
\begin{eqnarray}
\label{eq:sea13}
b & = & k~ b_{0} \nonumber \\
c & = & k~ c_{0}
\end{eqnarray}
Thus, in the limit of exact $SU(3)$ the $f/d$ ratio for the S-wave amplitudes
($(f/d)_{S-wave}~=~-1+(2c)/(3b)$)~ is the same as the ~$f_{0}/d_{0}$~ ratio for
the
$<B|H^{p.c.}_{weak}|B'>$ matrix elements
($f_{0}/d_{0}~=~-1+(2c_{0})/(3b_{0})$).

When $\delta s \ne 0$ from Eq.(\ref{eq:sea11}) we obtain
\begin{equation}
\label{eq:sea14}
A(\Sigma ^{+}_{0}) = \frac{1}{2\sqrt{2}} \frac{1}{1+x}~ k~ b_{0} -
\frac{1}{6\sqrt{2}} \frac{1}{1-x}~ k~ c_{0}
\end{equation}
where $x = \delta s/\Delta \omega _{s}$ .
All the other {\em pion}-emission amplitudes of Eq.(\ref{eq:pv2}) are modified
in the same way as $A(\Sigma ^{+}_{0})$, i.e.
\begin{eqnarray}
\label{eq:sea15}
k~ b_{0} \rightarrow & = & k~ b_{0}/(1+x) \nonumber \\
k~ c_{0} \rightarrow & = & k~ c_{0}/(1-x)
\end{eqnarray}
The above simple prescription does not apply to the {\em non-pion} emission
amplitudes which are, however, kinematically forbidden. Using $(f/d)_{S-wave} =
-1+ (2c)/(3b)$ and Eq.(\ref{eq:sea15}) one immediately obtains
Eq.(\ref{eq:in1}).

Inclusion of large quark sea effects explains the difference in the size of the
$(f/d)$ ratios of S- and P-wave amplitudes in a very natural way. At the same
time, however, the S-wave reduction mechanism proposed by LeYaouanc et al
(ref.\cite{LeY79}) to bring the S:P ratio into agreement with experiment
becomes essentially unimportant. Ref.\cite{LeY79} corresponds to $c_{0} = 0$
and leads to a reduction of S-wave amplitudes by 25-30\% for $x = 0.3$ to $0.4$
(see Eq.(\ref{eq:sea15}) and Table 5). With $f_{0}/d_{0} \approx -1.7$ ($c_{0}
\approx - b_{0}$) this reduction is, however, negligible. We shall discuss this
reappearing question of the S:P ratio in the next Section. Below, for
completeness and possible future use, we rewrite the $B_{i} \rightarrow
B_{f}P$   parity violating $\Delta S
=1$  amplitudes in an explicit $SU(3)$ language.

The relevant amplitudes are given by:
\newline
for $(b1)$ diagrams:
\begin{equation}
\label{eq:sea16a}
\frac{1}{2} Tr(SP^{\dagger}B^{\dagger}_{f}B_{i}) \frac{k b_{0}}{1-x}
\end{equation}
for $(b2)$ diagrams:
\begin{equation}
\label{eq:sea16b}
- \frac{1}{2} Tr(P^{\dagger}SB^{\dagger}_{f}B_{i}) \frac{k b_{0}}{1+x}
\end{equation}
for $(c1)$ diagrams
\begin{equation}
\label{eq:sea16c}
\frac{1}{6} [Tr(P^{\dagger}S[B^{\dagger}_{f},B_{i}]) - Tr(P^{\dagger}S)
Tr(B^{\dagger}_{f}B_{i})] \frac{k c_{0}}{1-x}
\end{equation}
for $(c2)$ diagrams
\begin{equation}
\label{eq:sea16d}
-\frac{1}{6} [Tr(SP^{\dagger}[B^{\dagger}_{f},B_{i}]) - Tr(P^{\dagger}S)
Tr(B^{\dagger}_{f}B_{i})] \frac{k c_{0}}{1+x}
\end{equation}
In Eqs.(\ref{eq:sea16a}-\ref{eq:sea16d})
\begin{eqnarray}
S & = & \lambda _{6} = \left( \begin{array}{ccc} 0 & 0 & 0 \\ 0 & 0 & 1 \\
0 & 1 & 0 \end{array}
\right)
\end{eqnarray}
is the spurion representing the weak Hamiltonian and $B_{i}$, $B_{f}$, $P$ are
the standard $3\times3$ matrices corresponding to the hadrons in question.

For the pions ($P=P_{\pi}$) only the $Tr(P^{\dagger}SB^{\dagger}_{f}B_{i})$ and
$Tr(P^{\dagger}SB_{i}B^{\dagger}_{f})$ traces in
Eqs.(\ref{eq:sea16a}-\ref{eq:sea16d}) are nonzero. Consequently, the
pion-emission amplitudes are:
\begin{eqnarray}
\label{eq:sea17a}
A(B_{i}\rightarrow B_{f}\pi) & = & -\frac{1}{2}
Tr(P^{\dagger}_{\pi}SB^{\dagger}_{f}B_{i}) \frac{k b_{0}}{1+x}
+ \frac{1}{6} Tr(P^{\dagger}_{\pi}S[B^{\dagger}_{f},B_{i}]) \frac{k c_{0}}{1-x}
\nonumber \\
 & = & d~ Tr(P^{\dagger}_{\pi}S\{ B_{i},B^{\dagger}_{f}\} ) + f~
Tr(P^{\dagger}_{\pi}S[B_{i},B^{\dagger}_{f}])
\end{eqnarray}
with
\begin{eqnarray}
\label{eq:sea17b}
d & = & -\frac{1}{4} \frac{k b_{0}}{1+x} \nonumber \\
f & = & \frac{1}{4} \frac{k b_{0}}{1+x} - \frac{1}{6} \frac{k c_{0}}{1-x}
\end{eqnarray}
and the apparent (i.e. applicable to pions amplitudes only) $(f/d)_{S-wave}$
ratio is given by
\begin{equation}
\label{eq:sea17c}
(f/d)_{S-wave} = -1+ \frac{2}{3} \frac{c_{0}}{b_{0}} \frac{1+x}{1-x}
\end{equation}

\section{The Problem of the S:P Ratio}
\label{S:P}
Large quark sea effects constitute an attractive explanation of the deviation
of $f/d$ from $-1$ because:

(1) their large size is consistent with unitarity-based calculations with the
scale provided by $\Delta-N$ splitting \cite{Zen92}, and

(2) they explain in a nice way the difference in the apparent $f/d$ ratios of
the S- and P-waves.

However, when sea effects are large, the S-wave reduction mechanism induced by
the $SU(3)$ breaking effects ceases to be significant and the problem of the
S:P ratio reappears. A possible way to deal with the latter has been discussed
by Milosevi\'c, Tadi\'c, and Trampeti\'c \cite{MTT82}, by Bonvin \cite{Bon84},
and by Nardulli \cite{Nar88}. These authors considered the radially excited
$(56,0^{+})^{*} \frac{1}{2}^{+}$ baryons $B^{*}$ in the intermediate states of
the P-wave amplitudes and found that their contribution has the same sign and
order of magnitude as the contribution from the ground-state baryons. The
details of the decomposition of the P-wave amplitudes into various
contributions differed in these papers substantially even though the
$f^{*}_{0}/d^{*}_{0}$ ratio for the radially excited baryons was assumed equal
to $-1$ in all these papers. Since in this paper we argue that for ground-state
baryons $f_{0}/d_{0}$ deviates from $-1$ significantly it is natural to expect
the same of $f^{*}_{0}/d^{*}_{0}$. In fact, it is natural to expect that
$f^{*}_{0}/d^{*}_{0} = f_{0}/d_{0}$: the relative size of the contributions
from the $W-$ exchange and sea pieces of the weak Hamiltonian in
$<B|H^{p.c.}_{weak}|B'>$ should be independent of whether the external state
$|B'>$ is a ground-state or radially excited baryon. The contributions from the
radially excited $\frac{1}{2} ^{+}$ baryons can be read off from the
weights of
Table~3. Assuming that radial excitations are heavier than the ground states by
$\Delta \omega ^{*} \approx 450~MeV$, the weights corresponding to diagrams
$(1)$ and $(2)$ of Fig.~2 have to be added leading to:
\begin{eqnarray}
\label{eq:sp18}
B(\Sigma ^{+}_{0}) & = &\frac{1}{\sqrt{2}} \left\{
\left[2 \left(1+\frac{F^{*}}{D^{*}} \right)
-
\left( 1-\frac{F^{*}}{D^{*}}\right) \right]
\left( 1-\frac{f^{*}_{0}}{d^{*}_{0}} \right)
\right\} G
\nonumber \\
B(\Sigma ^{+}_{+}) & = & \left\{ 2\left( 1+\frac{F^{*}}{D^{*}} \right)
\left( 1-\frac{f^{*}_{0}}{d^{*}_{0}} \right) - \frac{4}{3} \right\} G
\nonumber
\\ B(\Sigma ^{-}_{-}) & = & \left\{ \left( 1-\frac{F^{*}}{D^{*}} \right)
\left( 1-\frac{f^{*}_{0}}{d^{*}_{0}} \right) - \frac{4}{3} \right\}G  \\
B(\Lambda ^{0}_{-}) & = & \frac{1}{\sqrt{6}}
\left\{ -\left(1+\frac{F^{*}}{D^{*}}\right)
\left(3\frac{f^{*}_{0}}{d^{*}_{0}}+1\right)
+
2\left( 1-\frac{f^{*}_{0}}{d^{*}_{0}}\right) \right\} G \nonumber \\
B(\Xi ^{-}_{-})& = & \frac{1}{\sqrt{6}}
\left\{ 2\left( 1+\frac{f^{*}_{0}}{d^{*}_{0}}\right) +
\left( 1-\frac{F^{*}}{D^{*}}\right)
\left( 3\frac{f^{*}_{0}}{d^{*}_{0}}-1\right) \right\} G \nonumber
\end{eqnarray}
with
\begin{equation}
G = C \frac{\delta s}{\Delta \omega ^{*}} \frac{g^{*}}{g}
\end{equation}
In Eq.(\ref{eq:sp18})~ $F^{*}/D^{*}$ ($=0.56$) is the $F/D$ ratio for the
$B^{*}BP$ couplings and $g^{*}/g$ describes the relative size and sign of the
$B^{*}BP$ and $BBP$ couplings. (The ratio $g^{*}/g$ may be considered as
including the relative size of $d^{*}_{0}/d_{0}$, which in ref.\cite{Bon84} was
found to be close to 1, however.) In the quark model the ratio $g^{*}/g$ is
calculable and turns out to be {\em negative} and small (see e.g. Eq.(19) of
ref.\cite{Bon84}):
\begin{equation}
\label{eq:sp19}
\frac{g^{*}}{g} \approx -0.1~ {\rm to}~ -0.2
\end{equation}
In writing Eq.(\ref{eq:sp18}) we have neglected $SU(3)$ breaking in energy
denominators. Inclusion of this effect generates an additional contribution
whose symmetry structure is identical to that of the intermediate ground-state
baryons. It adds constructively to the latter one, though with a small relative
size of $-(\frac{\delta s}{\Delta \omega ^{*}})^{2} \frac{g^{*}}{g} \le 3\%$
only.

The contribution of the radially excited states (Eq.(\ref{eq:sp18})) violates
the Lee-Sugawara (LS) sum rule \cite{LeeSu}
\begin{equation}
\label{eq:sp20}
2 \Xi ^{-}_{-} + \Lambda ^{0}_{-} = \sqrt{3} \Sigma ^{+}_{0}
\end{equation}
for the P-waves. With the inclusion of radially excited states
Eq.(\ref{eq:sp20}) reads:
\begin{eqnarray}
\label{eq:sp21}
\frac{1}{\sqrt{6}} \left\{ 3\left( 1-\frac{f_{0}}{d_{0}} \right)
\left( \frac{F}{D}-1\right)
+
\frac{\delta
s}{\Delta \omega ^{*}} \frac{g ^{*}}{g}
3\left( 1+3\frac{F^{*}}{D^{*}}\right)
\left( 1-\frac{f^{*}_{0}}{d^{*}_{0}}\right)
\right\} C + & & \nonumber \\
+ ~\frac{1}{\sqrt{6}}
\frac{\delta
s}{\Delta \omega ^{*}} \frac{g ^{*}}{g} \left[
8\left( \frac{f^{*}_{0}}{d^{*}_{0}}
-
\frac{F^{*}}{D^{*}}\right)
\right] C
&
=
&
\nonumber
\\
= \frac{1}{\sqrt{6}} \left\{ 3\left(
1-\frac{f_{0}}{d_{0}}\right) \left( \frac{F}{D}-1 \right)
+
\frac{\delta s}{\Delta \omega ^{*}} \frac{g ^{*}}{g}
3\left( 1+3\frac{F^{*}}{D^{*}}\right) \left( 1-\frac{f^{*}_{0}}{d^{*}_{0}}
\right)
\right\} C
&
&
\nonumber
\\
 & &
\end{eqnarray}
Using experimental numbers Eq.(\ref{eq:sp20}) reads
\begin{equation}
\label{eq:sp22}
55.3 = 46.1
\end{equation}
The negative sign of $g*/g$ leads to the violation of LS rule in the direction
opposite to the experimental one. This violation comes about as follows. For
(all) $\Sigma$ and $\Lambda$ decays the contribution of radially excited states
has the same sign as that of ground-states and thus seems to help in the
explanation of the S:P ratio.However, for $\Xi$ decays this relative sign is
negative. If $ f^{*}_{0}/d^{*}_{0} = -1$ is used as in
refs.\cite{MTT82,Bon84,Nar88} the size of the contribution of radially excited
states to $\Xi$ decays is small. For $f^{*}_{0}/d^{*}_{0} = - 1.7$, however,
this contribution is bigger by a factor of 2.5 (see Eq.(\ref{eq:sp18})) and it
reduces the $\Xi ^{-}_{-}$ amplitudes (and the l.h.s. of Eq.(\ref{eq:sp20}))
very strongly. Consequently, only a small contribution (characterized by
$g^{*}/g \le - 0.05$) of the radially excited states can be tolerated if
$f^{*}_{0}/d^{*}_{0} \approx -1.7$. Inspection of Eqs.(\ref{eq:pc9}) and
(\ref{eq:sp18}) shows then that radially excited states may increase the
$\Lambda$, $\Sigma$ amplitudes by $\approx 15\%$ only. Thus, if
$f^{*}_{0}/d^{*}_{0} \approx -1.7$ the radially excited states cannot be held
responsible for the experimentally observed big size of P-wave amplitudes (or
small size of S-wave amplitudes).

In search for an explanation of the experimentally observed suppression of the
S:P ratio let us note that in the preceding sections we have pointed at
$SU(3)$ {\em symmetry breaking} as the possible origin of different deviations
of apparent
$f/d$ from $-1$. Thus, it was essentially proposed that the quark model as used
in the PCAC/CA approach has too much built-in symmetry. Similarly, the relative
size of various hadron couplings does not have to follow the quark model
predictions closely. For example, it is well known that the $\Delta \rightarrow
N$ magnetic transition is misjudged in the quark model by 30\% if the
magnetic moment of the proton is used to set the scale of quark-level
couplings. Now, $\Delta$ and $N$ are still members of the same $(56,0^{+})
\frac{1}{2} ^{+}$ $SU(6)\times O(3)$ multiplet. It is therefore conceivable
that similar or
bigger deviations from quark model predictions may appear when one attempts to
estimate the $B(\frac{1}{2} ^{+})B^{*}(\frac{1}{2} ^{-})P$ couplings from the
knowledge of familiar couplings of ground-state baryons to pseudoscalar
mesons. After all, we are dealing now with two different $SU(6)\times O(3)$
multiplets:
$(56,0^{+})$ and $(70,1^{-})$. A 30\% reduction in the overall size of the
$g_{B^{*}(1/2^{-})BP}$ and $<B|H^{p.v.}_{weak}|B^{*}(\frac{1}{2} ^{-})>$
couplings with respect to those calculated from $g_{BB'P}$ and
$<B|H^{p.c.}_{weak}|B'>$ by the quark-model route is a totally plausible
possibility. It would provide the missing factor of 2 by reducing $k$ in
Eq.(\ref{eq:sea12}) {\em without affecting} the relationship of
Eq.(\ref{eq:in1}) between the $f/d$ ratios of S- and P-wave amplitudes.
Clearly, the above argument constitutes a suggestion only. It would require a
thorough investigation which, for obvious reasons, is beyond the scope of this
paper: at the moment we do not know how to modify quark model to
improve its predictions for couplings.

\section{Weak Radiative Hyperon Decays}
\label{WRHD}
As already discussed in the preceding sections, in the literature on weak
nonleptonic hyperon decays there is no consensus on the origin of 1) the
suppression of the S:P ratio and 2) the deviations of $f/d$ from $-1$.
The general theoretical framework is not disputed, however. This is not the
case for
weak radiative hyperon decays which - for the last 25 years - have constituted
a real puzzle that has even been
termed "the last low-$q^{2}$ frontier of weak interaction
physics". For a thorough presentation of this highly controversial topic see
the recent review \cite{LaZen}. At present there is only one approach that
seems capable of describing fairly well the existing experimental data on
asymmetries and branching ratios of these decays. This approach, developed
recently by the author \cite{Zen89,Zen91}, is based on a combination of the
arguments of symmetry with the idea of vector meson dominance (VDM)
\cite{Sak69}. Although joint consideration of weak interactions, symmetry and
vector meson dominance looks innocent it is possible that it is intricately
linked with
very deep issues (see ref.\cite{LaZen}). Now, the $SU(3)$ symmetry breaking
effects discussed in the present paper have not been considered within that
approach as yet. Therefore it is of great importance to see how the results of
ref.\cite{Zen91} might be changed if $SU(3)$ symmetry breaking in energy
denominators is taken into account.

Calculation of the relevant weights is straightforward and leads to Table~6. In
this table only the weights corresponding to diagrams $(b1)$ and $(b2)$ have
been given. Contributions from diagrams $(c1)$ and $(c2)$ add up to the same
general $SU(3)$ structure irrespectively of whether $SU(3)$ is broken in energy
denominators or not. This general structure has been treated in
ref.\cite{Zen91} with the help of a parameter ($d'$ in Eq.(\ref{eq:sp18})
below). The $SU(3)$ symmetry breaking effects of the type considered in this
paper do affect the size of this parameter. However, they do not affect the
{\em relative} sizes of the single-quark contributions to various radiative
decays.
Since in the VDM $\times$ symmetry approach of ref.\cite{Zen91} $d'$ is treated
as a parameter to be fitted, $SU(3)$ symmetry breaking in energy denominators
is phenomenologically discernible in the contributions from the $(b)$ type
processes only.

The parity violating amplitudes due to $(b)$-type diagrams can be read off from
Table~5 and - together with the single-quark contributions - they give (up to
an overall VDM factor of $e/g$ ($e^{2}/(4\pi) = 1/137$, $g = 5.0$):
\begin{eqnarray}
\label{eq:wrhd61}
A(\Sigma ^{+} \rightarrow p\gamma) & = &
- \frac{b}{9\sqrt{2}} \left\{ 2 + \epsilon + 3 \frac{1+x}{1-x} \right\} +
\frac{1}{\sqrt{2}}~ d' \nonumber \\
A(\Sigma ^{0} \rightarrow n\gamma) & = &
- \frac{b}{18} \left\{ 3 \frac{1+x}{1-x} - 2 - \epsilon \right\} - \frac{1}{2}~
d'
\nonumber \\
A(\Lambda \rightarrow n\gamma)    & = &
\frac{b}{6\sqrt{3}} \left\{ 2 + \epsilon + \frac{1+x}{1-x} \right\} -
\frac{3\sqrt{3}}{2}
{}~d'  \\
A(\Xi ^{0} \rightarrow \Lambda \gamma) & = &
- \frac{2+\epsilon}{9\sqrt{3}}~ b + \frac{\sqrt{3}}{2}~ d'\nonumber\\
A(\Xi ^{0} \rightarrow \Sigma ^{0} \gamma) & = &
- \frac{1}{3}~ b~ \frac{1+x}{1-x} - \frac{5}{2}~ d' \nonumber \\
A(\Xi ^{-} \rightarrow \Sigma ^{-} \gamma) & = &
\frac{5}{\sqrt{2}}~ d' \nonumber
\end{eqnarray}
with $ b = k b_{0}/(1+x) = -5$ (in units of $10^{-7}$) and small negative $d'$.

{}From Eq.(\ref{eq:wrhd61}) it follows that - when compared to the SU(3)
symmetric case ($x=0$) - the $SU(3)$ symmetry breaking in energy denominators:

(1) increases the parity violating amplitudes in the $\Sigma ^{+} \rightarrow
p\gamma$, $\Sigma ^{0} \rightarrow n\gamma$, $\Lambda \rightarrow n\gamma$ and
$\Xi ^{0} \rightarrow \Sigma ^{0} \gamma$ decays

(2) leaves the $\Xi ^{0} \rightarrow \Lambda \gamma$ parity violating amplitude
unchanged.

No change of sign of the $(b)$-type two-quark contribution is observed. Since
the contribution of the single-quark parity violating amplitudes (terms
proportional to $d'$ in Eq.(\ref{eq:wrhd61})) is strongly limited from above by
the recently measured branching ratio of the $\Xi ^{-} \rightarrow \Sigma ^{-}
\gamma$ decay \cite{Dub93}, we conclude that the basic expectations
of the $SU(3)$ symmetric approach of ref.\cite{Zen91} (such as signs and
approximate size of asymmetries) cannot change much when the effect of $SU(3)$
symmetry breaking in the denominators is included. However, the slight increase
in the value of $A(\Lambda \rightarrow n\gamma)$ would make it easier to fit
the observed $\Lambda \rightarrow n\gamma$ branching ratio \cite{Nob92}. At the
same time, the increase of $A(\Xi ^{0} \rightarrow \Sigma ^{0} \gamma)$ would
manifest itself mostly in a more negative asymmetry of the $\Xi ^{0}
\rightarrow \Sigma ^{0} \gamma$ decay. The only experiment performed so far
\cite{Tei89} yields a slightly positive (albeit with a large error) value for
this asymmetry ($+0.2\pm0.32$).
The calculations of this paper confirm therefore that it is very important to
measure the $\Xi ^{0} \rightarrow \Sigma ^{0} \gamma$ asymmetry precisely.
Should this asymmetry stay significantly positive it would add yet another
question mark to the long-standing enigma of weak radiative hyperon decays.

\section{Summary}
\label{summary}
In this paper we have carried out an analysis of the joint influence of large
quark sea and $SU(3)$-symmetry breaking effects in weak hyperon decays. An
explanation of the difference between the values of the apparent $f/d$ ratios
for the S- and P-wave amplitudes of nonleptonic decays has been proposed. It
was pointed out that quark sea effects in the matrix elements of the parity
conserving part of the weak Hamiltonian between the ground-state baryons are
additionally enhanced in the S-wave amplitudes by the presence of the
$SU(3)$-symmetry breaking effects in energy denominators. A formula for this
enhancement has been derived and shown to agree with the data extremely well if
the dominant part of the deviation of $(f/d)_{P-wave}$ from $-1$ is due to sea
quarks. This corroborates our earlier calculations which indicated that large
deviations of $(f/d)_{soft~ pion}$ from its naive quark model value of $-1$ are
to be expected when the quark model is properly unitarized. Thus,
the commonly used quark model value of $(f/d)_{soft~ pion} = -1$
should be replaced by a value close to -1.7. We suggest that a possible
way to resolve the S:P problem is to break the naive quark model predictions
relating the values of matrix elements involving the $(56,0^{+})$ and
$(70,1^{-})$ baryons. In view of unsolved difficulties existing elsewhere in
similar problems involving baryon couplings this possible route of explaining
the S:P size problem cannot be properly handled at the moment: we do not know
how to modify the (oversimplified) naive quark model predictions for couplings.
Finally, implications of this paper for the weak radiative hyperon decays have
been briefly discussed. It was shown that the signs of
the asymmetries previously calculated in the $SU(3)$-symmetric approach are
unchanged by the inclusion of the
$SU(3)$-symmetry breaking effects in energy denominators.



\renewcommand{\baselinestretch}{2}
\small
\normalsize
\newpage
\baselineskip =10pt
\begin{center}
\begin{tabular}{|l|l|r|r|r|r|}
\hline
 & transition & $(b1)$ & $(b2)$ & $(c1)$ & $(c2)$ \\ \hline
$\Sigma ^{+}_{0}$ & $\Sigma ^{+} \rightarrow p \pi ^{0}$ & $ 0$ &
$\frac{1}{2\sqrt{2}}$ & $ - \frac{1}{6\sqrt{2}}$ & $0$ \\
$\Sigma ^{+}_{+}$ & $\Sigma ^{+} \rightarrow n \pi ^{+}$ & $ 0$ & $ 0$& $0$ &
$0$ \\
 & $\Sigma ^{+} \rightarrow p \eta_{8}$ & $-\frac{1}{\sqrt{6}}$ &
$-\frac{1}{2\sqrt{6}}$ & $\frac{1}{6\sqrt{6}}$ & $\frac{1}{3\sqrt{6}}$ \\
$\Sigma ^{-}_{-}$ & $\Sigma ^{-} \rightarrow n \pi ^{-}$ & $0$ & $-\frac{1}{2}$
& $\frac{1}{6}$ & $0$ \\
\hline
$\Lambda ^{0}_{-}$ & $\Lambda \rightarrow p \pi ^{-}$ & $0$ &
$-\frac{1}{2\sqrt{6}}$ & $\frac{1}{2\sqrt{6}}$ & $0$ \\
$\Lambda ^{0}_{0}$ & $\Lambda \rightarrow n \pi ^{0}$ & $0$ &
$\frac{1}{4\sqrt{3}}$ & $-\frac{1}{4\sqrt{3}}$ & $0$ \\
 & $\Lambda \rightarrow n \eta _{8}$ & $-\frac{1}{6}$ & $-\frac{1}{12}$&
$\frac{1}{12}$&
$\frac{1}{6}$ \\ \hline
$\Xi ^{-}_{-}$ & $\Xi ^{-} \rightarrow \Lambda \pi ^{-}$ & $0$ &
$\frac{1}{\sqrt{6}}$ & $-\frac{1}{2\sqrt{6}}$ & $0$ \\
$\Xi ^{0}_{0}$ & $\Xi ^{0} \rightarrow \Lambda \pi^{0}$ & $0$ &
$-\frac{1}{2\sqrt{3}}$ & $\frac{1}{4\sqrt{3}}$ & $0$ \\

 & $\Xi ^{0} \rightarrow \Lambda \eta_{8}$ & $\frac{1}{3}$ &
$\frac{1}{6}$ & $-\frac{1}{12}$ & $-\frac{1}{6}$  \\
\hline
 & $\Xi ^{-} \rightarrow \Sigma ^{-} \pi ^{0}$ & $0$ & $0$ &
$\frac{1}{6\sqrt{2}}$ & $0$ \\
 & $\Xi ^{-} \rightarrow \Sigma ^{-} \eta _{8}$ & $0$ & $0$ &
$-\frac{1}{6\sqrt{6}}$ & $-\frac{1}{3\sqrt{6}}$ \\
\hline
 & $p \rightarrow K^{0}p$ & $0$ & $-\frac{1}{2}$ & $0$ & $\frac{1}{6}$ \\
\hline
\end{tabular}
\end{center}
\vskip 1cm
\baselineskip = 10pt

Table 1. Weights of quark diagrams $(b)$ and $(c)$ for the S-wave (p.v.)
amplitudes.
\newpage
\baselineskip = 20pt
\begin{center}
\begin{tabular}{|l|l|r|r|r|r|}
\hline
 & transition & $(b1)$ & $(b2)$ & $(c1)$ & $(c2)$ \\
\hline
$\Sigma ^{+}_{0}$ & $\Sigma ^{+} \rightarrow p \pi ^{0}$ & $
-\frac{1}{3\sqrt{2}}$
&
$-\frac{1}{6\sqrt{2}}$ & $ - \frac{1}{9\sqrt{2}}$ & $0$ \\
$\Sigma ^{+}_{+}$ & $\Sigma ^{+} \rightarrow n \pi ^{+}$ & $ -\frac{1}{3}$ & $
0$&
$0$
&
$0$ \\
 & $\Sigma ^{+} \rightarrow p \eta_{8}$ & $0$ &
$-\frac{1}{2\sqrt{6}}$ & $\frac{1}{9\sqrt{6}}$ & $-\frac{2}{9\sqrt{6}}$ \\
$\Sigma ^{-}_{-}$ & $\Sigma ^{-} \rightarrow n \pi ^{-}$ & $0$ & $\frac{1}{6}$
& $\frac{1}{9}$ & $0$ \\
\hline
$\Lambda ^{0}_{-}$ & $\Lambda \rightarrow p \pi ^{-}$ & $-\frac{1}{3\sqrt{6}}$
&
$-\frac{1}{2\sqrt{6}}$ & $-\frac{1}{\sqrt{6}}$ & $0$ \\
$\Lambda ^{0}_{0}$ & $\Lambda \rightarrow n \pi ^{0}$ & $\frac{1}{6\sqrt{3}}$ &
$\frac{1}{4\sqrt{3}}$ & $\frac{1}{2\sqrt{3}}$ & $0$ \\
 & $\Lambda \rightarrow n \eta _{8}$ & $0$ & $\frac{1}{4}$&
$-\frac{1}{6}$&
$\frac{1}{3}$ \\ \hline
$\Xi ^{-}_{-}$ & $\Xi ^{-} \rightarrow \Lambda \pi ^{-}$ & $0$ &
$\frac{1}{3\sqrt{6}}$ & $\frac{1}{3\sqrt{6}}$ & $0$ \\
$\Xi ^{0}_{0}$ & $\Xi ^{0} \rightarrow \Lambda \pi^{0}$ & $0$ &
$-\frac{1}{6\sqrt{3}}$ & $-\frac{1}{6\sqrt{3}}$ & $0$ \\

 & $\Xi ^{0} \rightarrow \Lambda \eta_{8}$ & $0$ &
$-\frac{1}{6}$ & $\frac{1}{18}$ & $-\frac{1}{9}$  \\
\hline
 & $\Xi ^{-} \rightarrow \Sigma ^{-} \pi ^{0}$ & $0$ & $0$ &
$-\frac{5}{9\sqrt{2}}$ & $0$ \\
 & $\Xi ^{-} \rightarrow \Sigma ^{-} \eta _{8}$ & $0$ & $0$ &
$\frac{5}{9\sqrt{6}}$ & $-\frac{10}{9\sqrt{6}}$ \\
\hline
 & $p \rightarrow K^{0}p$ & $0$ & $-\frac{1}{6}$ & $0$ & $-\frac{1}{9}$ \\
\hline
\end{tabular}
\end{center}
\vskip 1cm
\baselineskip = 5pt

Table 2. Weights of quark diagrams $(b)$ and $(c)$ for the P-wave (p.c.)
amplitudes.
\newpage
\baselineskip =10pt
\begin{center}
\begin{tabular}{|l|c|c|}
\hline
 & Fig.2.1 & Fig.2.2 \\
\hline
$\Sigma ^{+}_{0}$ & $\frac{1}{\sqrt{2}} \left(1+\frac{F}{D}\right) \left(
1-\frac{f}{d} \right)$ & $\frac{1}{\sqrt{2}} \left(1+\frac{F}{D}\right) \left(
1-\frac{f}{d} \right) - \frac{1}{\sqrt{2}} \left(1-\frac{F}{D}\right) \left(
1-\frac{f}{d} \right)$ \\
$\Sigma ^{+}_{+}$ & $\left(1+\frac{F}{D}\right) \left(1-\frac{f}{d} \right)$
& $\left(1+\frac{F}{D}\right) \left(1-\frac{f}{d} \right) - \frac{4}{3}$ \\
$\Sigma ^{-}_{-}$ & 0 &
$\left(1-\frac{F}{D}\right) \left(1-\frac{f}{d} \right) - \frac{4}{3}$ \\
\hline
$\Lambda ^{0}_{-}$ & $-\frac{1}{\sqrt{6}} \left( 1+\frac{F}{D} \right)
\left( 1+3\frac{f}{d} \right)$ & $\frac{2}{\sqrt{6}} \left( 1-\frac{f}{d}
\right)$ \\
\hline
$\Xi ^{-}_{-}$ & $\frac{2}{\sqrt{6}} \left( 1+\frac{f}{d} \right)$ &
$\frac{1}{\sqrt{6}} \left( 1- \frac{F}{D} \right) \left( 3 \frac{f}{d} -1
\right)$ \\
\hline
\end{tabular}
\end{center}
\vskip 0.7 cm
\baselineskip =5pt

Table 3. Weights of baryon pole contributions for the P-wave amplitudes.
\baselineskip =10pt
\vskip 1.6cm

\begin{center}
\begin{tabular}{|c|c|c|c|c|c|}
\hline
 & \multicolumn{4}{c|}{F/D=0.56,~C=-33} & \\
\cline{2-5}
process & \multicolumn{3}{c|}{baryon legs} & meson leg & data \\
\cline{2-5}
        & $f_{0}/d_{0}=-1.7$ & $-1.85$ & $-1.9$ & $f'_{0}/d_{0}=-0.15$ & \\
\hline
$\Sigma ^{+}_{0}$ & 27.7 & 29.3 & 29.8 & 1.6 & $26.6 \pm 1.3$ \\
$\Sigma ^{+}_{+}$ & 44.0 & 44.0 & 44.0 & 0 &   $42.4 \pm 0.35$ \\
$\Sigma ^{-}_{-}$ & 4.8 & 2.6 & 1.9 & -2.2 &   $-1.44 \pm 0.17$ \\
$\Lambda ^{0}_{-}$& 13.4 & 18.8 & 20.6 & 5.4 & $22.1 \pm 0.5$ \\
$\Xi ^{-}_{-}$ & 17.3 & 15.9 & 15.5 & -1.4 & $16.6 \pm 0.8 $ \\
\hline
\end{tabular}
\end{center}
\vskip 0.7cm

Table 4. P-wave amplitudes (in units of $10^{-7}$) from Eq.(\ref{eq:pc9}).

\newpage
\begin{center}
\begin{tabular}{|c|c|c|c|c|c|}
\hline
& \multicolumn{2}{c|}{$x=0$} & \multicolumn{2}{c|}{$x=1/3$} & data \\
$f_{0}/d_{0}$ & -1 & -1.7 & -1 & -1.7 & \\
\hline
$\Sigma ^{+}_{0}$ & -5.4 & -7.3 & -4.1 & -6.8 & -3.27 \\
$\Sigma ^{+}_{+}$ & 0    &   0  &  0   &   0  & 0.13 \\
$\Sigma ^{-}_{-}$ & 7.7  & 10.3 & 5.8  & 9.7  & 4.27 \\
$\Lambda ^{0}_{-}$& 3.1  & 6.4  & 2.3  & 7.1  & 3.23  \\
$\Xi ^{-}_{-}$    & -6.3 & -9.5 & -4.7 & -9.5 & -4.50 \\
\hline
\end{tabular}
\end{center}
\vskip 0.4cm
\baselineskip = 5pt

Table 5. S-wave amplitudes (in units of $10^{-7}$) as calculated from P-wave
amplitudes
\baselineskip =10 pt
\vskip 1.4cm

\begin{center}
\begin{tabular}{|l|r|r|}
\hline
transition & $(b1)$ & $(b2)$ \\
\hline
$\Sigma ^{+} \rightarrow p \gamma$ & $-\frac{1}{3\sqrt{2}}$ &
$-\frac{2+\epsilon}{9\sqrt{2}}$ \\
$\Sigma ^{0} \rightarrow n \gamma$ & $-\frac{1}{6}$ & $\frac{2+\epsilon}{18}$
\\
$\Lambda \rightarrow n \gamma$ & $\frac{1}{6\sqrt{3}}$ &
$\frac{2+\epsilon}{6\sqrt{3}}$ \\
$\Xi ^{0} \rightarrow \Lambda \gamma$ & $0$ & $-\frac{2+\epsilon}{9\sqrt{3}}$
\\
$\Xi ^{0} \rightarrow \Sigma ^{0} \gamma$ & $\frac{1}{3}$ & $0$ \\
$\Xi ^{-} \rightarrow \Sigma ^{-} \gamma$ & $0$ &$0$ \\
\hline
\end{tabular}
\end{center}
\vskip 0.4 cm

\baselineskip =5pt
Table 6. Weights of quark diagrams $(b)$ for the radiative S-wave amplitudes.
\baselineskip = 10pt

\newpage
\setlength{\unitlength}{0.55pt}
\begin{picture}(550,1000)
\put(60,99){
\begin{picture}(450,900)
\put(0,720){
\begin{picture}(220,160)
\put(100,15){\makebox(0,0){(a)}}
\put(10,65){\makebox(0,0){$B_{f}$}}
\put(190,65){\makebox(0,0){$B_{i}$}}
\put(70,140){\makebox(0,0){$M$}}
\put(30,90){\line(1,0){25}}
\put(85,90){\vector(-1,0){30}}
\put(115,90){\line(1,0){30}}
\put(170,90){\vector(-1,0){25}}
\put(85,150){\vector(0,-1){30}}
\put(85,120){\line(0,-1){30}}
\put(115,90){\vector(0,1){30}}
\put(115,120){\line(0,1){30}}
\put(170,65){\vector(-1,0){70}}
\put(100,65){\line(-1,0){70}}
\put(170,40){\vector(-1,0){70}}
\put(100,40){\line(-1,0){70}}
\multiput(85,90)(5,0){6}{\line(1,0){3}}
\end{picture}}
\put(250,720){
\begin{picture}(200,160)
\put(100,15){\makebox(0,0){(a')}}
\put(170,65){\vector(-1,0){70}}
\put(100,65){\line(-1,0){70}}
\put(170,40){\vector(-1,0){70}}
\put(100,40){\line(-1,0){70}}
\put(85,150){\vector(0,-1){20}}
\put(115,150){\line(0,-1){20}}
\put(115,130){\vector(0,1){0}}
\put(100,130){\oval(30,30)[b]}
\put(170,90){\vector(-1,0){25}}
\put(145,90){\vector(-1,0){90}}
\put(55,90){\line(-1,0){25}}
\multiput(100,90)(0,5){5}{\line(0,1){3}}
\end{picture}}
\put(0,180){
\begin{picture}(200,160)
\put(100,15){\makebox(0,0){(d1)}}
\put(30,90){\line(1,0){25}}
\put(85,90){\vector(-1,0){30}}
\put(115,90){\line(1,0){30}}
\put(170,90){\vector(-1,0){25}}
\put(85,150){\vector(0,-1){30}}
\put(85,120){\line(0,-1){30}}
\put(115,90){\vector(0,1){30}}
\put(115,120){\line(0,1){30}}
\put(170,65){\vector(-1,0){70}}
\put(100,65){\line(-1,0){70}}
\put(170,40){\vector(-1,0){70}}
\put(100,40){\line(-1,0){70}}
\multiput(130,40)(0,5){5}{\line(0,1){3}}
\end{picture}}
\put(250,180){
\begin{picture}(200,160)
\put(100,15){\makebox(0,0){(d2)}}
\put(30,90){\line(1,0){25}}
\put(85,90){\vector(-1,0){30}}
\put(115,90){\line(1,0){30}}
\put(170,90){\vector(-1,0){25}}
\put(85,150){\vector(0,-1){30}}
\put(85,120){\line(0,-1){30}}
\put(115,90){\vector(0,1){30}}
\put(115,120){\line(0,1){30}}
\put(170,65){\vector(-1,0){70}}
\put(100,65){\line(-1,0){70}}
\put(170,40){\vector(-1,0){70}}
\put(100,40){\line(-1,0){70}}
\multiput(70,40)(0,5){5}{\line(0,1){3}}
\end{picture}}

\put(0,540){
\begin{picture}(200,160)
\put(100,15){\makebox(0,0){(b1)}}
\put(170,90){\vector(-1,0){25}}
\put(145,90){\line(-1,0){50}}
\multiput(130,65)(0,5){5}{\line(0,1){3}}
\put(65,90){\vector(-1,0){20}}
\put(45,90){\line(-1,0){15}}
\put(65,150){\vector(0,-1){30}}
\put(65,120){\line(0,-1){30}}
\put(95,90){\vector(0,1){30}}
\put(95,120){\line(0,1){30}}
\put(170,65){\vector(-1,0){90}}
\put(80,65){\line(-1,0){50}}
\put(170,40){\vector(-1,0){90}}
\put(80,40){\line(-1,0){50}}
\end{picture}}
\put(0,360){
\begin{picture}(200,160)
\put(100,15){\makebox(0,0){(c1)}}
\put(170,90){\vector(-1,0){15}}
\put(155,90){\line(-1,0){10}}
\put(115,90){\line(-1,0){20}}
\multiput(115,90)(5,0){6}{\line(1,0){3}}
\put(130,90){\oval(30,30)[b]}
\put(130,75){\vector(-1,0){0}}
\put(65,90){\vector(-1,0){20}}
\put(45,90){\line(-1,0){15}}
\put(65,150){\vector(0,-1){30}}
\put(65,120){\line(0,-1){30}}
\put(95,90){\vector(0,1){30}}
\put(95,120){\line(0,1){30}}
\put(170,65){\vector(-1,0){90}}
\put(80,65){\line(-1,0){50}}
\put(170,40){\vector(-1,0){90}}
\put(80,40){\line(-1,0){50}}
\end{picture}}
\put(250,540){
\begin{picture}(200,160)
\put(100,15){\makebox(0,0){(b2)}}
\put(170,90){\vector(-1,0){15}}
\put(155,90){\line(-1,0){20}}
\put(135,90){\vector(0,1){30}}
\put(135,120){\line(0,1){30}}
\put(105,150){\vector(0,-1){30}}
\put(105,120){\line(0,-1){30}}
\put(170,65){\vector(-1,0){50}}
\put(120,65){\line(-1,0){90}}
\put(170,40){\vector(-1,0){50}}
\put(120,40){\line(-1,0){90}}
\put(105,90){\vector(-1,0){50}}
\put(55,90){\line(-1,0){25}}
\multiput(70,65)(0,5){5}{\line(0,1){3}}
\end{picture}}
\put(250,360){
\begin{picture}(200,160)
\put(100,15){\makebox(0,0){(c2)}}
\put(170,90){\vector(-1,0){15}}
\put(155,90){\line(-1,0){20}}
\put(135,90){\vector(0,1){30}}
\put(135,120){\line(0,1){30}}
\put(105,150){\vector(0,-1){30}}
\put(105,120){\line(0,-1){30}}
\put(170,65){\vector(-1,0){50}}
\put(120,65){\line(-1,0){90}}
\put(170,40){\vector(-1,0){50}}
\put(120,40){\line(-1,0){90}}
\put(105,90){\line(-1,0){20}}
\multiput(55,90)(5,0){6}{\line(1,0){3}}

\put(55,90){\vector(-1,0){10}}
\put(45,90){\line(-1,0){15}}
\put(70,90){\oval(30,30)[b]}
\put(70,75){\vector(-1,0){0}}
\end{picture}}
\put(0,0){
\begin{picture}(200,160)
\put(100,15){\makebox(0,0){(e1)}}
\put(170,65){\vector(-1,0){15}}
\put(155,65){\line(-1,0){10}}
\put(115,65){\line(-1,0){35}}

\multiput(115,65)(5,0){6}{\line(1,0){3}}
\put(130,65){\oval(30,30)[b]}

\put(130,50){\vector(-1,0){0}}
\put(65,90){\vector(-1,0){20}}
\put(45,90){\line(-1,0){15}}
\put(65,150){\vector(0,-1){30}}
\put(65,120){\line(0,-1){30}}
\put(95,90){\vector(0,1){30}}
\put(95,120){\line(0,1){30}}
\put(170,90){\vector(-1,0){75}}
\put(80,65){\line(-1,0){50}}
\put(170,40){\vector(-1,0){90}}
\put(80,40){\line(-1,0){50}}
\end{picture}}
\put(250,0){
\begin{picture}(200,160)
\put(100,15){\makebox(0,0){(e2)}}
\put(170,90){\vector(-1,0){15}}
\put(155,90){\line(-1,0){20}}
\put(135,90){\vector(0,1){30}}
\put(135,120){\line(0,1){30}}
\put(105,150){\vector(0,-1){30}}
\put(105,120){\line(0,-1){30}}
\put(170,65){\vector(-1,0){50}}
\put(105,90){\line(-1,0){75}}
\put(170,40){\vector(-1,0){50}}
\put(120,40){\line(-1,0){90}}
\put(120,65){\line(-1,0){35}}
\multiput(55,65)(5,0){6}{\line(1,0){3}}
\put(55,65){\vector(-1,0){10}}
\put(45,65){\line(-1,0){15}}
\put(70,65){\oval(30,30)[b]}
\put(70,50){\vector(-1,0){0}}
\end{picture}}

\end{picture}}
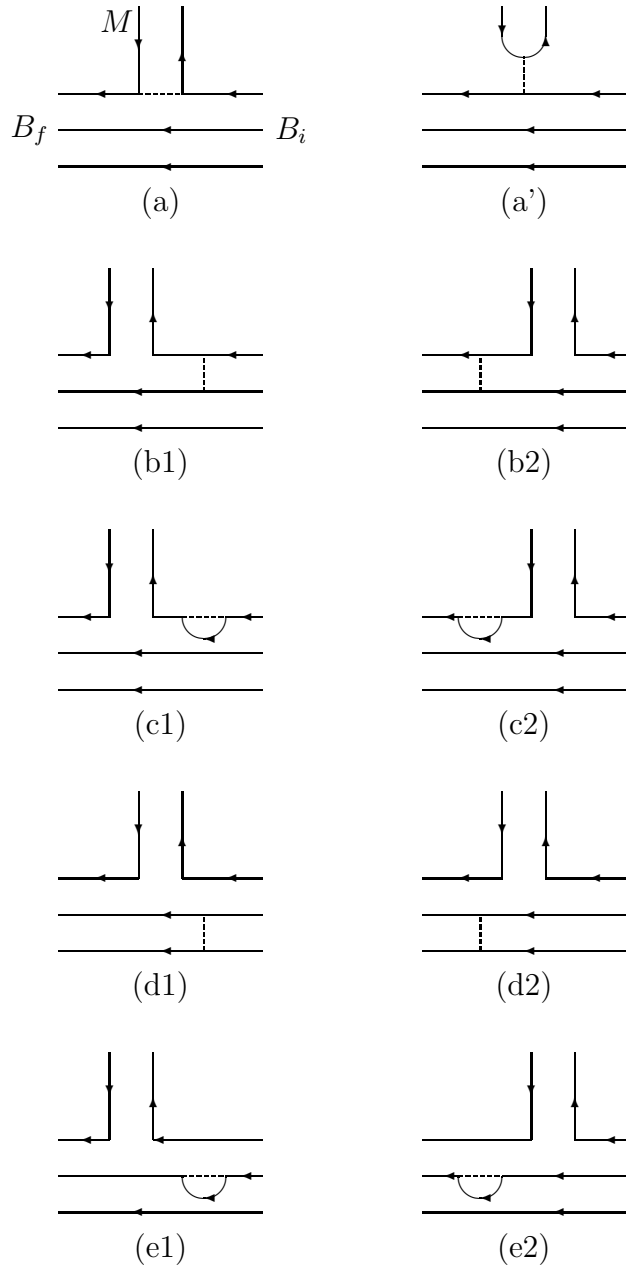
\put(300,10){\makebox(0,0)[b]{Fig.1. Quark diagrams for weak decays.}}
\end{picture}
\newpage
\begin{picture}(600,330)
\put(0,100){
\begin{picture}(250,220)
\put(30,80){\line(-1,0){30}}
\put(70,80){\vector(-1,0){40}}
\put(125,80){\line(-1,0){55}}
\put(180,80){\vector(-1,0){55}}
\put(220,80){\line(-1,0){40}}
\put(250,80){\vector(-1,0){30}}
\multiput(70,80)(0,6){15}{\line(0,1){3}}
\put(180,80){\circle*{10}}
\put(180,50){\makebox(0,0)[b]{$H_{weak}$}}
\put(220,90){\makebox(0,0)[b]{$B_{i}$}}
\put(120,90){\makebox(0,0)[b]{$B$, $B^{*}$}}
\put(30,90){\makebox(0,0)[b]{$B_{f}$}}
\put(90,150){\makebox(0,0)[b]{$M$}}
\put(125,0){\makebox(0,0)[b]{$(1)$}}
\end{picture}
}
\put(350,100){
\begin{picture}(250,220)
\put(30,80){\line(-1,0){30}}
\put(70,80){\vector(-1,0){40}}
\put(125,80){\line(-1,0){55}}
\put(180,80){\vector(-1,0){55}}
\put(220,80){\line(-1,0){40}}
\put(250,80){\vector(-1,0){30}}
\multiput(180,80)(0,6){15}{\line(0,1){3}}
\put(70,80){\circle*{10}}
\put(70,50){\makebox(0,0)[b]{$H_{weak}$}}
\put(220,90){\makebox(0,0)[b]{$B_{i}$}}
\put(120,90){\makebox(0,0)[b]{$B$, $B^{*}$}}
\put(30,90){\makebox(0,0)[b]{$B_{f}$}}
\put(200,150){\makebox(0,0)[b]{$M$}}
\put(125,0){\makebox(0,0)[b]{$(2)$}}

\end{picture}
}
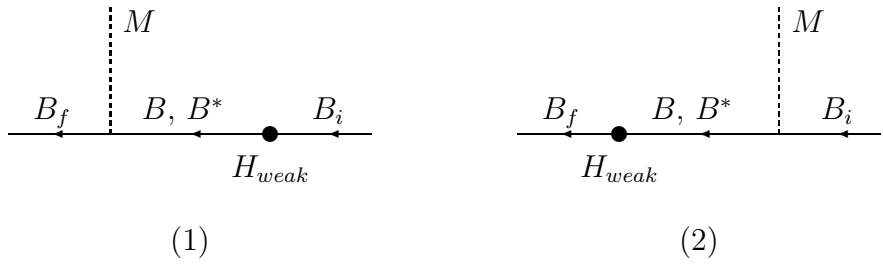
\put(300,10){\makebox(0,0)[b]{Fig.2. Baryon-pole diagrams for weak decays.}}
\end{picture}
\end{document}